\documentclass[12pt]{article}
\usepackage{array}
\begin{document}
\newcommand{\bq}{\begin{equation}}
\newcommand{\eq}{\end{equation}}
\newcommand{\bqa}{\begin{eqnarray}}
\newcommand{\eqa}{\end{eqnarray}}
\newcommand{\nl}{\nonumber \\}
\newcommand{\eqn}[1]{Eq.(\ref{#1})}
\newcommand{\Tab}[1]{Tab.\ref{#1}}
\newcommand{\from}{\leftarrow}
\newcommand{\Exp}{\mathsf{E}}
\newcommand{\cpi}[1]{\cos\left({\pi #1\over2}\right)}
\newcommand{\spi}[1]{\sin\left({\pi #1\over2}\right)}
\newcommand{\tcpu}{t_{\rm{cpu}}}
\newcommand{\ndim}{m}
\newcommand{\ouralg}{{\tt OURALG}}
\newcommand{\reject}{{\tt REJECT}}

\title{{\bf A fast algorithm for generating a uniform distribution
            inside a high-dimensional polytope}}
\author{Andr\'e van Hameren\thanks{{\tt andrevh@sci.kun.nl}}~ and 
        Ronald Kleiss\thanks{{\tt kleiss@sci.kun.nl}}\\
	University of Nijmegen, Nijmegen, the Netherlands}
\maketitle	
\begin{abstract}
We describe a uniformly fast algorithm for generating
points $\vec{x}$ uniformly in a hypercube with the restriction that
the difference between each pair of coordinates is bounded.
We discuss the quality of the algorithm in the sense of its usage of
pseudo-random source numbers, and present an interesting result on the
correlation between the coordinates.
\end{abstract}

\thispagestyle{empty}
\newpage
\pagestyle{plain}
\setcounter{page}{1}

\section{Introduction}
In this paper we shall discuss the problem of generating sets of
points $\vec{x}=(x_1,x_2,\ldots,x_m)$
inside an $m$-dimensional hypercube with an additional
restriction. The points $\vec{x}$ are required to satisfy the
conditions
\bq
|x_k| < 1\;\;\;,\;\;\; |x_k-x_l| < 1\;\;\mbox{for all $k,l$}\;\;.
\label{polytopedefinition}
\eq
These conditions define a $m$-dimensional convex polytope $P$.
The reason for tackling this problem is the following. In a recently
developed Monte Carlo algorithm, {\tt SARGE} \cite{sarge}, we
address the problem of generating configurations of four-momenta
$p_i^\mu$, $i=1,2,\ldots,n$ of $n$ massless partons at high energy,
with a distribution that has, as much as possible, the form of a so-called
QCD antenna:
$$
{1\over s_{12}s_{23}s_{34}\cdots s_{n-1,n}s_{n1}}\;\;\;,
\;\;\;s_{kl}=(p_k+p_l)^2\;\;,
$$
where $s_{kl}$ is the invariant mass squared of partons $k$ and $l$,
with the additional requirement that the total invariant mass squared of
all the partons is fixed to $s$, and {\em every\/} $s_{kl}$ (also those
not occurring explicitly in the antenna) exceeds some lower bound $s_0$:
in this way the singularities of the QCD matrix elements are avoided.
The {\tt SARGE} algorithm has a structure that is, in part, similar
to the {\tt RAMBO} algorithm \cite{rambo},
where generated momenta are scaled so as to
attain the correct overall invariant mass. Obviously, in {\tt SARGE} this is
more problematic because of the $s_0$ cut, but one should like to implement
this cut as far as possible. Note that out of the $n(n-1)/2$ different
$s_{kl}$, $n$ occur in the antenna, and each of these must of course
be bounded by $s_0$ from below and some $s_M < s$ from above. The
scale-invariant ratios of two of these masses are therefore bounded by
\bq
{s_0\over s_M} \le {s_{ij}\over s_{kl}} \le {s_M\over s_0}\;\;,
\eq
The structure of the {\tt SARGE} algorithm is such \cite{sarge} that
there are $m=2n-4$ of these ratios to be generated. By going over to
variables 
$$
x_{(\cdots)}=\log(s_{ij}/s_{kl})/\log(s_M/s_0)\;\;,
$$ 
and inspecting
all ratios that can be formed from the chosen $m$ ones, we arrive at
the condition of \eqn{polytopedefinition}. Note that, inside {\tt SARGE},
a lot of internal rejection is going on, and events satisfying
\eqn{polytopedefinition} {\em may\/} still be discarded: however, if
\eqn{polytopedefinition} is not satisfied, the event is {\em certainly\/}
discarded, and it therefore pays to include this condition from the start.

\section{The algorithm}
The most straightforward way of implementing is of course the following:
generate $x_k$, $k=1,\ldots,m$ by $x_k \from 2\rho-1$, and reject
if the conditions are not met.
Here and in the following, each occurrence of $\rho$ stands for a call to
a source of iid uniform pseudo-random numbers between in $[0,1)$.
The drawback of this approach is
that the efficiency, {\it i.e.} the probability of success per try,
is given by $2^{-m}V_m(P)$ (where $V_m(P)$ is the volume of the polytope $P$)
and becomes very small for large $m$, as we shall see.

To compute the volume $V_m(P)$ we first realize that the condition
$|x_k-x_l|<1$ is only relevant when $x_k$ and $x_l$ have opposite sign.
Therefore, we can divide the $x$ variables in 
$m-k$ positive and $k$ negative ones, so that
\bqa
V_{m,k}(P) &=& \int\limits_0^1
dy_1 dy_2 \cdots dy_k dx_{k+1} dx_{k+2}\cdots dx_m
\theta\left(1-\max_ix_i - \max_jy_j\right)\;\;,\nl
V_m(P) &=& \sum\limits_{k=0}^m {m!\over k!(m-k)!}V_k(P)\;\;,
\eqa
where we have written $y_k=-x_k$. By symmetry we can always relabel the
indices such that $x_m = \max_ix_i$ and $y_1 = \max_jy_j$. The integrals
over the other $x$'s and $y$'s can then easily be done, and we find
\bqa
V_{m,k}(P) &=& k(m-k)\int\limits_0^1 dy_1 y_1^{k-1}
\int\limits_0^{1-y_1} dx_m x_m^{m-k-1} \nl &=&
k\int\limits_0^1 dy_1 y_1^{k-1}(1-y_1)^{m-k} = {k!(m-k)!\over m!}\;\;,
\eqa
and hence
\bq
V_m(P) = m+1\;\;.
\eq
The efficiency of the straightforward algorithm is therefore
equal to $(m+1)/2^m$, which is less than 3\% for $n$ larger than 6.

We have given the above derivation explicitly since it allows us, by
working backwards, to find a rejection-free algorithm with unit efficiency.
The algorithm is as follows:
\begin{enumerate}
\item Choose a value for $k$. Since each $k$ is exactly equally probably
we simply have
$$
k \from \lfloor (m+1)\rho \rfloor\;\;.
$$
\item For $k=0$ we can simply put
$$
x_i \from \rho\;\;\;,i=1,\ldots,m\;\;,
$$
while for $k=m$ we put
$$
x_i \from -\rho\;\;\;,i=1,\ldots,m\;\;. 
$$
\item
For $0<k<m$, $y_1$ has the unnormalized density $y_1^{k-1}(1-y_1)^{m-k}$
  between 0 and 1.
      An efficient algorithm to do this is Cheng's rejection algorithm BA for
      beta random variates (cf. \cite{Devroye})\footnote{There is an error on
      page 438 of \cite{Devroye}, where ``$V\from\lambda^{-1}U_1(1-U_1)^{-1}$''
      should be replaced by ``$V\from\lambda^{-1}\log[U_1(1-U_1)^{-1}]$''.},
      but the following also works:
      $$
        v_1\from -\log\left(\prod_{i=1}^k\rho\right)\;\;,\;\;
        v_2\from -\log\left(\prod_{j=1}^{m-k+1}\rho\right)\;\;,\;\;
        y_1 \from \frac{v_1}{ v_1+v_2}\;\;.
      $$
The variable $x_m$ has unnormalized density $x_m^{m-k-1}$ between 0 and $1-y_1$
so that it is generated by
$$
x_m \from (1-y_1)\rho^{1/(m-k)}\;\;.
$$
The other $x$'s are now trivial:
\bqa
x_1 \from -y_1\;\;&,&\;\;x_i \from x_1\rho,\;\;i=2,3,\ldots,k\;\;,\nl
& &\;\;x_i \from x_m\rho,\;\;i=k+1,k+2,\ldots,m-1\;\;.\nonumber
\eqa
Finally, perform a random permutation of the whole set 
$(x_1,x_2,\ldots,x_m)$.
\end{enumerate}

\section{Computational complexity}
The number usage $S$, that is, the
expected number of calls to the random number source 
$\rho$ per event can be derived
easily. In the first place, 1 number is used to get $k$ for every event.
In a fraction $2/(m+1)$ of the cases, only $m$ calls are made.
In the remaining cases, there are $k+(m-k+1) = m+1$ calls to get
$y_1$, and 1 call for all the other $x$ values.
Finally, the simplest permutation algorithm calls $m-1$ times \cite{knuth}.
The expected number of calls is therefore
\begin{equation}
   S =1 + \frac{2m}{ m+1} + \frac{m-1}{m+1}(m+1 + (m-1) + (m-1)) =
   {3m^2 - m + 2\over m+1}\;\;.
\end{equation}
For large $m$ this comes to about $3m-1$ calls per event.
Using a more sophisticated permutation algorithm would use at least 1 call,
giving
\begin{equation}
  S = 1 + \frac{2m}{ m+1} + \frac{m-1}{m+1}(m+1 + (m-1) + (1)) = 2m\;\;.
\end{equation}
We observed that Cheng's rejection algorithm to obtain $y_1$ uses about
2 calls per event. Denoting this number by $C$ the expected number
of calls becomes
\bq
S = {2m^2 + (C-1)m - C + 3\over m+1} \sim 2m + C - 1
\eq
for the simple permutation algorithm, while the more sophisticated one
would yield
\bq
S = {m^2 + (C+2)m - C +1\over m+1} \sim m + C + 2\;\;.
\eq
We see that in all these cases the algorithm is uniformly efficient
in the sense that the needed number of calls is simply
proportional to the problem's complexity $m$, as $m$ becomes large.
An ideal algorithm would of course still need $m$ calls, while the
straightforward rejection algorithm rather has
$S = m2^m/(m+1) \sim 2^m$ expected calls per event.

In the testing of algorithms such as this one, it is useful to study
expectation values of, and correlations between, the various $x_i$.
Inserting either $x_i$ or $x_ix_j$ in the integral expression for
$V(P)$, we found after some algebra the following expectation
values:
\begin{equation}
 \Exp(x_i) = 0\;\;\;,\;\;\;
 \Exp(x_i^2) = \frac{m+3}{6(m+1)}\;\;\;,\;\;\;
 \Exp(x_ix_j) = \frac{m+3}{12(m+1)}\;\;(i\ne j)\;\;,
\end{equation}
so that the correlation coefficient 
between two different $x$'s is precisely 1/2
in all dimensions! This somewhat surprising fact allows for a simple
but powerful check on the correctness of the algorithm's implementation.

As an extra illustration of the efficiency, we present in the tables below the
cpu-time ($\tcpu$) needed to generate $1000$ points in an $\ndim$-dimensional
polytope, both with the algorithm presented in this paper (\ouralg) and the
rejection method (\reject). In the latter, we just
\begin{enumerate}
\item put $x_i\from 2\rho-1$ for $i=1,\ldots,\ndim$;
\item reject $\vec{x}$ if $|x_i-x_j|>1$ for 
      $i=1,\ldots,\ndim-1$ and $j=i+1,\ldots,\ndim$.
\end{enumerate}
The computations were done using a single $333$-MHz UltraSPARC-IIi processor,
and the random number generator used was {\tt RANLUX} on level 3.
\[
\begin{array}{|c|c|c|}
  \hline
  \multicolumn{1}{|c|}{}&\multicolumn{2}{|c|}{\tcpu ({\rm sec})}\\\hline
  \ndim & \ouralg & \reject\\ \hline
    2 & 0.03  & 0.01  \\ \hline
    3 & 0.03  & 0.02  \\ \hline
    4 & 0.03  & 0.04  \\ \hline
    5 & 0.04  & 0.08  \\ \hline
    6 & 0.05  & 0.17  \\ \hline
    7 & 0.06  & 0.32  \\ \hline
    8 & 0.07  & 0.67  \\ \hline
    9 & 0.08  & 1.33  \\ \hline
   10 & 0.09  & 2.76  \\ \hline
\end{array}
\qquad
\begin{array}{|c|c|r|}
  \hline
  \ndim & \ouralg & \reject\\ \hline
   11 & 0.09  &    5.15 \\ \hline
   12 & 0.10  &   10.94 \\ \hline
   13 & 0.11  &   21.71 \\ \hline
   14 & 0.12  &   44.06 \\ \hline
   15 & 0.13  &   87.90 \\ \hline
   16 & 0.14  &  169.65 \\ \hline
   17 & 0.15  &  336.67 \\ \hline
   18 & 0.16  &  671.46 \\ \hline
   19 & 0.17  & 1383.33 \\ \hline
   20 & 0.18  & 2744.82 \\ \hline
\end{array}
\]
For $\ndim=2$ and $\ndim=3$, the rejection method is quicker, but from
$\ndim=4$ on, the cpu-time clearly grows linearly for the method presented in 
this paper, and exponentialy for the rejection method.

\section{Extension}
Let us, finally, comment on one possible extension of this algorithm. 
Suppose that the points $\vec{x}$ are distributed on the polytope $P$,
but with an additional (unnormalized) density given by
\bq
F(\vec{x}) = \prod\limits_{i=1}^m\cpi{x_i}\;\;,
\eq
so that the density is suppressed near the edges. It is then still
possible to compute $V_{m,k}(P)$ for this new density:
\bqa
V_{k,m}(P) &=& k(m-k)\int\limits_0^1 dy_1\cpi{y_1}
\int\limits_0^{1-y_1}dx_m\cpi{x_m}\nl
& &\left(\int\limits_0^{y_1}dy\cpi{y}\right)^{k-1}
\left(\int\limits_0^{x_m}dx\cpi{x}\right)^{m-k-1}\nl
&=& k(m-k)\left({2\over\pi}\right)^m
\int\limits_0^1d\spi{y_1}\left(\spi{y_1}\right)^{k-1}\nl
& &\int\limits_0^{\cpi{y_1}}d\spi{x_m}\left(\spi{x_m}\right)^{m-k-1}\nl
&=& {2^{m-1}k\over\pi^m}\int\limits_0^1
ds\;s^{k/2-1}(1-s)^{(m-k)/2}\nl
&=& \left({2\over\pi}\right)^m{\Gamma(1+k/2)\Gamma(1+(m-k)/2)\over
\Gamma(1+m/2)}\;\;,
\eqa
where we used $s=\left(\spi{y_1}\right)^2$.
Therefore, a uniformly efficient algorithm can be constructed in this
case as well, along the following lines.
Using the $V_{k,m}$, the relative weights for each $k$ can
be determined. Then $s$ is generated as a $\beta$ distribution.
The generation of the other $x$'s involves only manipulations with sine
and arcsine functions. Note that, for large $m$, the weighted volume
of the polytope $P$ is
\bqa
V(P) &=& \sum\limits_{k=0}^m
\left({2\over\pi}\right)^m
{\left({k\over2}\right)!\left({m-k\over2}\right)!\over\left({m\over2}\right)!}
{m!\over k!(m-k)!} \nl
&\sim& m\sqrt{{\pi\over8}}\left({8\over\pi^2}\right)^{m/2}\;\;,
\eqa
so that a straightforward rejection algorithm would have number usage
\bq
S \sim \sqrt{{8\over\pi}}\left({\pi^2\over2}\right)^{m/2}\;\;,
\eq
and a correspondingly decreasing efficiency.

\end{document}